# Optical Phonons of SnSe$_{(1-x)}$S$_x$ Layered Semiconductor Alloys

Tharith Sriv[1,2], Thi Minh Hai Nguyen[3], Yangjin Lee[4,5], Soo Yeon Lim[1], Van Quang Nguyen[3], Kwanpyo Kim[4,5], Sunglae Cho[3] and Hyeonsik Cheong[1,*]

[1]Department of Physics, Sogang University, Seoul 04107, Korea
[2]Department of Physics, Royal University of Phnom Penh, Phnom Penh, Cambodia
[3]Department of Physics and Energy Harvest Storage Research Center (EHSRC), University of Ulsan, Ulsan 44610, Korea
[4]Department of Physics, Yonsei University, Seoul 03722, Korea
[5]Center for Nanomedicine, Institute for Basic Science (IBS), Seoul 03722, Korea
[*]Coresponding author: hcheong@sogang.ac.kr

**ABSTRACT**

The evolution of the optical phonons in layered semiconductor alloys SnSe$_{(1-x)}$S$_x$ is studied as a function of the composition by using polarized Raman spectroscopy with six different excitation wavelengths (784.8, 632.8, 532, 514.5, 488, and 441.6 nm). The polarization dependences of the phonon modes are compared with transmission electron diffraction measurements to determine the crystallographic orientation of the samples. Some of the Raman modes show significant variation in their polarization behavior depending on the excitation wavelengths. It is established that the maximum intensity direction of the A$_g^2$ mode of SnSe$_{(1-x)}$S$_x$ (0≤$x$≤1) does not depend on the excitation wavelength and corresponds to the armchair direction. It is additionally found that the lower-frequency Raman modes of A$_g^1$, A$_g^2$ and B$_{3g}^1$ in the alloys show the typical one-mode behavior of optical phonons, whereas the higher-frequency modes of B$_{3g}^2$, A$_g^3$ and A$_g^4$ show two-mode behavior.

**Introduction**

Interest in graphene and other two-dimensional (2D) materials such as MoS$_2$ and black phosphorus (BP) has continuously increased since the first isolation of graphene in 2004[1]. Tin (II) selenide (SnSe), tin (II) sulfide (SnS) and their alloys are an interesting class of 2D materials for a variety of reasons: their band gap falls between 0.9 eV (SnSe) and 1.5 eV (SnS)[2-4], an important range for photovoltaic and optoelectronic applications[5-13]; due to the in-plane anisotropy of the crystal, the optical and electronic properties depend on the in-plane direction of a device; and unlike a similar material, black phosphorus, they are fairly stable in air[13-15]. Furthermore, a very high thermoelectric figure of merit (ZT) value of 2.6 was reported in bulk SnSe, which showed dependence on the crystallographic direction[16]. Therefore, determination of the crystallographic direction is crucial in utilizing these materials in device applications. Polarized Raman spectroscopy is a tool of choice for determining the crystallographic direction of anisotropic materials[14, 17-24]. There have been a few reports on polarized Raman spectroscopy of SnS, SnSe, and SnSe$_{0.5}$S$_{0.5}$[14, 21-22, 25-26], and unpolarized Raman measurements have been reported for a few more compositions[26]. However, a systematic study over the full alloy composition is still lacking. More importantly, previous studies did not take into account the excitation wavelength dependence of the polarized Raman signals: because the polarization dependence of the Raman signals in anisotropic 2D materials sometimes depend sensitively on the excitation wavelength[17], polarized Raman spectroscopy has to be compared with a direct determination of the crystallographic direction using electron microscopy.

In this work, a systematic investigation of the polarization dependence of Raman modes of temperature gradient grown[27-29] SnSe$_{(1-x)}$S$_x$ single crystal alloys ($x$=0, 0.2, 0.4, 0.5, 0.6, 0.7, 0.8, 0.9, 1) was conducted by using low-frequency polarized Raman spectroscopy in room temperature with six excitation wavelengths. The evolution of the phonon modes as a function of the alloy composition was analyzed in detail. It was found that of the six main phonon modes, the three lower frequency modes show the one-mode behavior, whereas the higher frequency modes exhibit the two-mode behavior. By comparing the polarization dependence of the phonon modes with the transmission electron microscopy (TEM) measurements, the crystallographic direction was determined without any ambiguity. It was found that the $A_g^2$ mode, for example, at 70 cm$^{-1}$ for SnSe or at 95 cm$^{-1}$ for SnS, or the corresponding mode in the alloys, is the most reliable mode for determination of the crystallographic direction regardless of the excitation wavelength.

**Results and Discussion**

Figure 1(a) shows the crystal structure of SnSe and SnS, where the zigzag and armchair directions are chosen along the *y*- and *z*-axes, respectively, and the exfoliated surfaces are (100) planes. The lattice constants at room temperature are *a*=11.501 Å, *b*=4.153 Å and *c*=4.445 Å for SnSe and *a*=11.14 Å, *b*=3.97 Å and *c*=4.34 Å for SnS, respectively[2, 4]. Figure 1(b) shows the x-ray diffraction (XRD) patterns of the SnSe$_{(1-x)}$S$_x$ (0, 0.2, 0.4, 0.5, 0.6, 0.8, 1) single crystals, and Figure 1(c) gives the lattice constant *a* for each composition extracted from the XRD patterns and a comparison with the estimated value of *a* obtained from calculations using Vegard's law. The calculated values of *b* and *c* are also shown in the same figure. The peak positions (2$\theta$) correlate well with the nominal compositions, and the sharp peaks indicate high crystal quality (see Supplementary Figure S1and S2 for EDS results on all the samples studied). The estimated compositions do not deviate from the nominal values significantly.

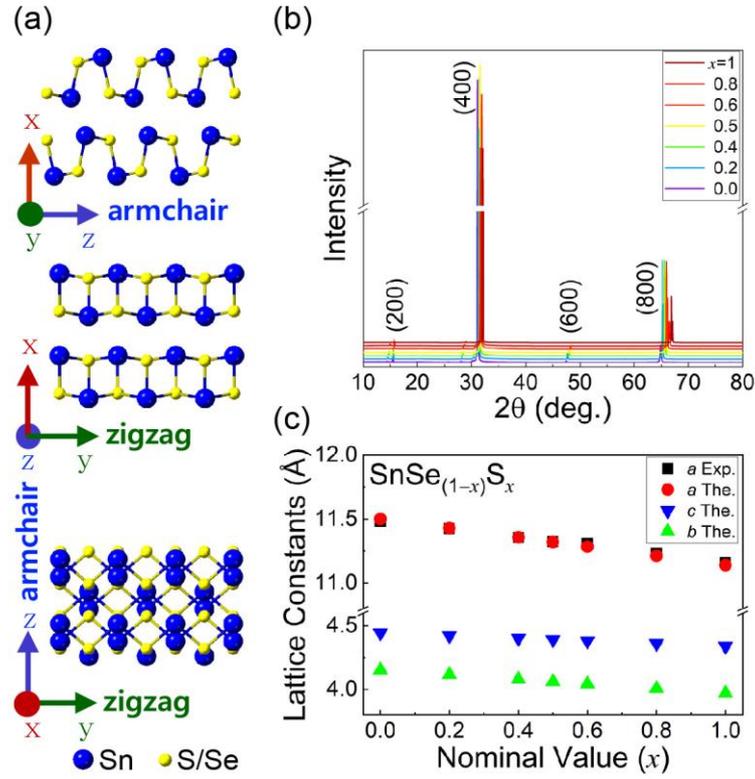

**Figure 1**. (**a**) Crystal structure of SnS and SnSe. (**b**) XRD patterns of the SnSe$_{(1-x)}$S$_x$ ($x$=0, 0.2, 0.4, 0.5, 0.6, 0.8, 1) single crystals. (**c**) Lattice constants as a function of nominal composition ($x$). The lattice constant $a$ extracted from the XRD patterns marked as Exp. is compared with that from theoretical (The.) calculation (Vegard's law) as indicated. The lattice constants $b$ and $c$ are from theoretical calculations.

SnSe and SnS have the orthorhombic structure of $D_{2h}^{16}$ (*Pnma*) space group with eight atoms per unit cell[30-31]. There exist 24 vibrational modes in the Brillouin zone center, which can be represented by, $\Gamma = 4A_g + 2B_{1g} + 4B_{2g} + 2B_{3g} + 2A_u + 4B_{1u} + 2B_{2u} + 4B_{3u}$. Among 21 optical phonons, 2 are inactive, 7 infrared active, and 12 Raman active. The polarization dependence of the Raman intensity of each Raman active mode can be described by $|\hat{e}_i \cdot R \cdot \hat{e}_s^T|^2$, where $\hat{e}_i$ and $\hat{e}_s$ represent the polarizations of the incident and scattered radiations at the position of scattering, and $R$ is the Raman tensor being expressed as[21-22, 25-26],

$$A_\text{g} = \begin{pmatrix} A & 0 & 0 \\ 0 & B & 0 \\ 0 & 0 & C \end{pmatrix}, B_{1\text{g}} = \begin{pmatrix} 0 & D & 0 \\ D & 0 & 0 \\ 0 & 0 & 0 \end{pmatrix}, B_{2\text{g}} = \begin{pmatrix} 0 & 0 & E \\ 0 & 0 & 0 \\ E & 0 & 0 \end{pmatrix}, \text{and } B_{3\text{g}} = \begin{pmatrix} 0 & 0 & 0 \\ 0 & 0 & F \\ 0 & F & 0 \end{pmatrix}.$$

The $B_{1\text{g}}$ and $B_{2\text{g}}$ modes are forbidden in the backscattering geometry[32-33] with the incident laser beam in the *x*-direction. Therefore, six Raman modes should be observed in the Raman spectra of pure compounds SnSe or SnS. For the SnSe$_{(1-x)}$S$_x$ alloys, this analysis is only approximately valid in the virtual crystal approximation: as it will be shown below, there are more than six modes in the alloys.

In anisotropic layered materials such as black phosphorus or ReS$_2$, it is well known that the polarization dependences of the phonon modes vary with the excitation wavelength[17, 19, 34]. This peculiar behavior is ascribed to the birefringence and linear dichroism as well as the anisotropy of the electron-phonon interaction in this class of materials, which can be described in terms of the dielectric tensor and its dependence on the wavelength. By writing the Raman tensor elements, which is related to the derivative of the polarizability in the semi-classical model, in terms of complex numbers, the variation of the polarization dependences as a function of the excitation wavelength has been successfully described[17, 19, 34]. In our analysis, we write the Raman tensor elements as complex numbers: $A = |A|e^{i\phi_A}, B = |B|e^{i\phi_B}, C = |C|e^{i\phi_C}, D = |D|e^{i\phi_D}, E = |E|e^{i\phi_E}$, and $F = |F|e^{i\phi_F}$. In backscattering geometry with the light traveling in the *x*-direction, the polarization vectors can be written as $\hat{e}_i = \hat{e}_s = (0, \cos\theta, \sin\theta)$ for parallel (∥) and $\hat{e}_i = (0, \cos\theta, \sin\theta)$, $\hat{e}_s = (0, \cos(\theta + 90°), \sin(\theta + 90°)) = (0, -\sin\theta, \cos\theta)$ for perpendicular (⊥, cross) polarization configurations, respectively. This results in the intensity of the observed modes ($A_\text{g}$ and $B_{3\text{g}}$) to be expressed as,

$$I(A_\text{g}, \parallel) \propto |B|^2 \cos^4\theta + |C|^2 \sin^4\theta + 2|B||C|\sin^2\theta\cos^2\theta\cos\phi_{CB}, \quad (1)$$

$$I(B_{3\text{g}}, \parallel) \propto |F|^2 \sin^2 2\theta, \quad (2)$$

$$I(A_g, \perp) \propto \frac{1}{4} \times (|B|^2 + |C|^2 - 2|B||C|\cos\phi_{CB})\sin^2 2\theta, \text{ and} \quad (3)$$

$$I(B_{3g}, \perp) \propto |F|^2 \cos^2 2\theta, \tag{4}$$

where $\emptyset_{CB} = \emptyset_C - \emptyset_B$ is the phase difference of the Raman tensor elements.

Figure 2(a-c) show the polarization dependence of Raman spectra of SnSe, SnSe$_{0.5}$S$_{0.5}$ and SnS measured by using the 632.8-nm excitation wavelength in parallel configuration. The excitation wavelength dependence of the Raman spectra was examined (see Supplementary Figure S3, S4, S5). Most peaks show maximum intensity with the 514.5-nm excitation. However, the lowest-frequency B$_{3g}$ mode is clearly resolved only for 632.8- and 785-nm excitations. Therefore, the 632.8-nm excitation wavelength was chosen for most of the analysis. For SnSe (SnS), six modes are observed as expected: A$_g$ modes at 32, 70, 131 and 150 cm$^{-1}$ (40, 95, 191, 219 cm$^{-1}$) and B$_{3g}$ at 37 and 109 cm$^{-1}$ (49 and 164 cm$^{-1}$), which are in agreement with previous studies on pure compounds[21-22, 25-26]. For alloys, more than six peaks are resolved as shown in Figure 2(b) (see Supplementary Figure S6 for the other compositions studied). Figure 2(d) is an expanded plot of the Raman spectrum of SnSe$_{0.5}$S$_{0.5}$ in the range of 100 to 250 cm$^{-1}$. In addition to the strong peaks at 121 cm$^{-1}$ (B$_{3g}$), 140 cm$^{-1}$ (A$_g$), and 199 cm$^{-1}$ (A$_g$), two more weak peaks are required in the deconvolution to reproduce the experimental spectra using Lorentzian functions. Since a signal centered at 154 cm$^{-1}$ has a different polarization dependence from that of the stronger peak at about 140 cm$^{-1}$, it can be identified by comparing the polarization dependences. The much weaker signal at 186 cm$^{-1}$ is identified through careful examination of different data sets. This is a SnS-like B$_{3g}$ mode as will be explained later. Its position corresponds to the extrapolated position of the SnS-like B$_{3g}$ mode from the data for higher *x* values (see Supplementary Figure S7 for deconvolution procedure and results for the other compositions). Figure 2(e) also shows that the signal at this frequency has a polarization dependence like that of the peak at 121 cm$^{-1}$ and is clearly different from that of the signal at 199 cm$^{-1}$. These assignments were confirmed by polarized Raman measurements in the cross-polarization

configuration (see Supplementary Figure S8, S9). Similar analyses were carried out for all the alloy compositions (Figure S6, S7). The polarization dependences of the identified Raman modes are consistent with their mode assignments and the expected polarization dependences as summarized in Supplementary Figures S10 and S11.

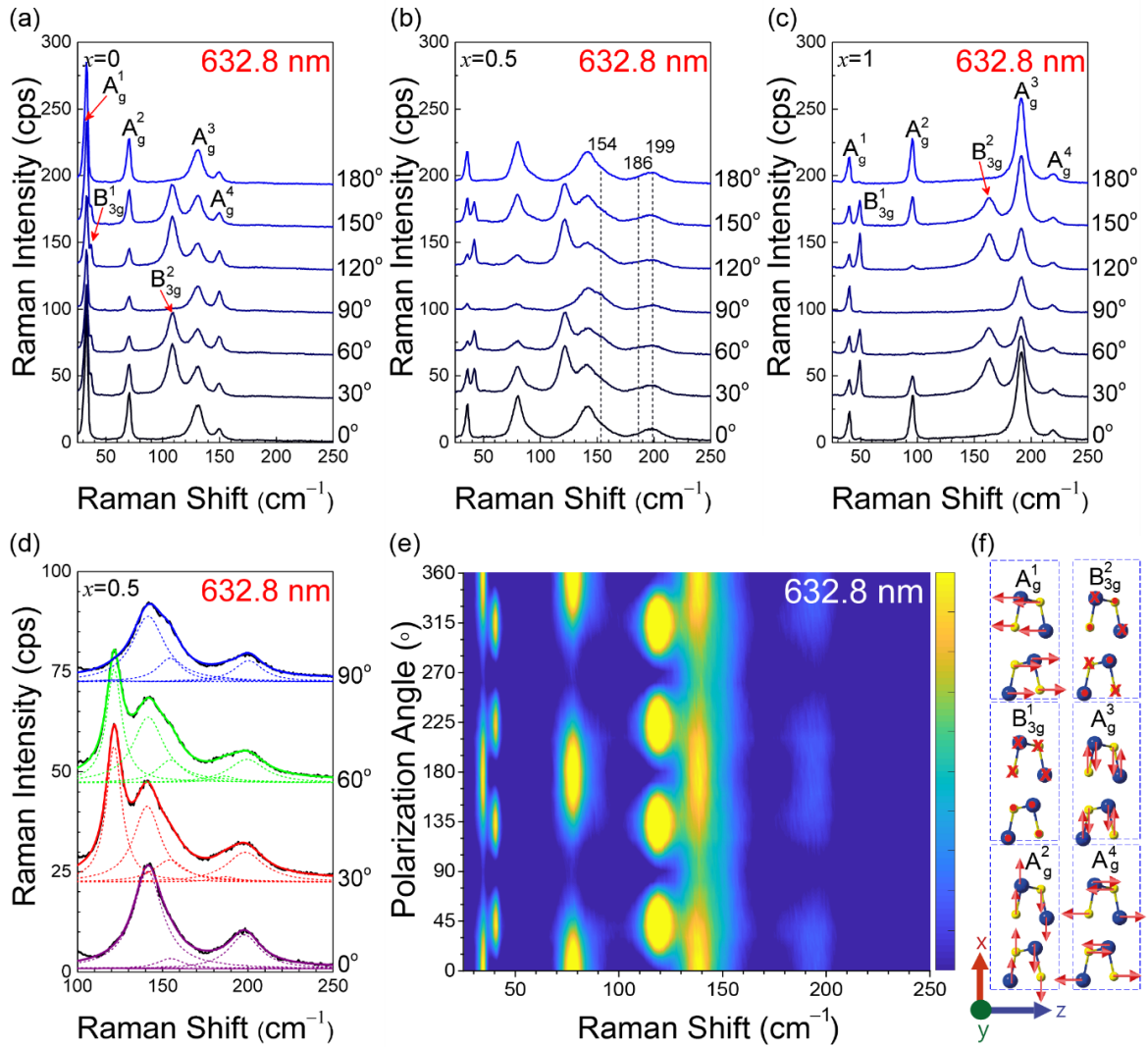

**Figure 2**. (**a-c**) Polarization dependence of the Raman spectra of SnSe$_{(1-x)}$S$_x$ ($x$=0, 0.5, 1) with the 632.8-nm excitation in parallel polarization configuration. The mode assignments for the pure compounds ($x$=0, 1) are shown. Gray vertical dotted lines in (**b**) indicate the position of weak peaks in SnSe$_{0.5}$S$_{0.5}$. (**d**) Deconvolution of the Raman spectrum of the SnSe$_{0.5}$S$_{0.5}$ alloy in the range of 100 cm$^{-1}$ to 250 cm$^{-1}$ for different polarizations. (**e**) Polarization dependence of Raman spectrum intensity of SnSe$_{0.5}$S$_{0.5}$ alloy, measured in 10°

increments. (**f**) Vibrational modes corresponding to the observed Raman modes in SnSe and SnS[22].

Figure 3(a) shows the dependence of the Raman spectrum in the 20 cm$^{-1}$-300 cm$^{-1}$ band on the composition $x$. Here, the polarizations of the incident and scattered photons are set to 30° to show all the peaks with reasonable intensities. Figure 3(d) shows the spectra in the 100 cm$^{-1}$-250 cm$^{-1}$ with the fitting functions. Individual peaks were deconvoluted by using multiple Lorentzian functions (see Supplementary Figure S7 for more fitting results). Figure 3(b) summarizes the Raman peak positions measured in several different polarization configurations. The shade of the symbols corresponds to the normalized intensity of each Raman mode. Interesting trends are observed. The three Raman peaks in the low-frequency region (<100 cm$^{-1}$) shifts monotonically with $x$ for the entire composition range. This is a so-called 'one mode behavior' of phonon modes in alloys. On the other hand, the peaks in the higher frequency range show different behavior. As the composition $x$ is increased from 0 (SnSe), the peaks shift to higher frequency, which is similar to the case of low-frequency modes. However, a new set of peaks appear at higher frequencies as the composition $x$ is further increased. As the composition approaches $x=1$, this new set of peaks become dominant, whereas the peaks that evolve from the Raman peaks of SnSe ($x=0$) fade away. This is a typical 'two-mode behavior' in alloys[35]. The peaks that evolve from the SnSe phonon peaks correspond to 'SnSe-like' modes, whereas the peaks that become stronger for S-rich compositions correspond to 'SnS-like' modes.

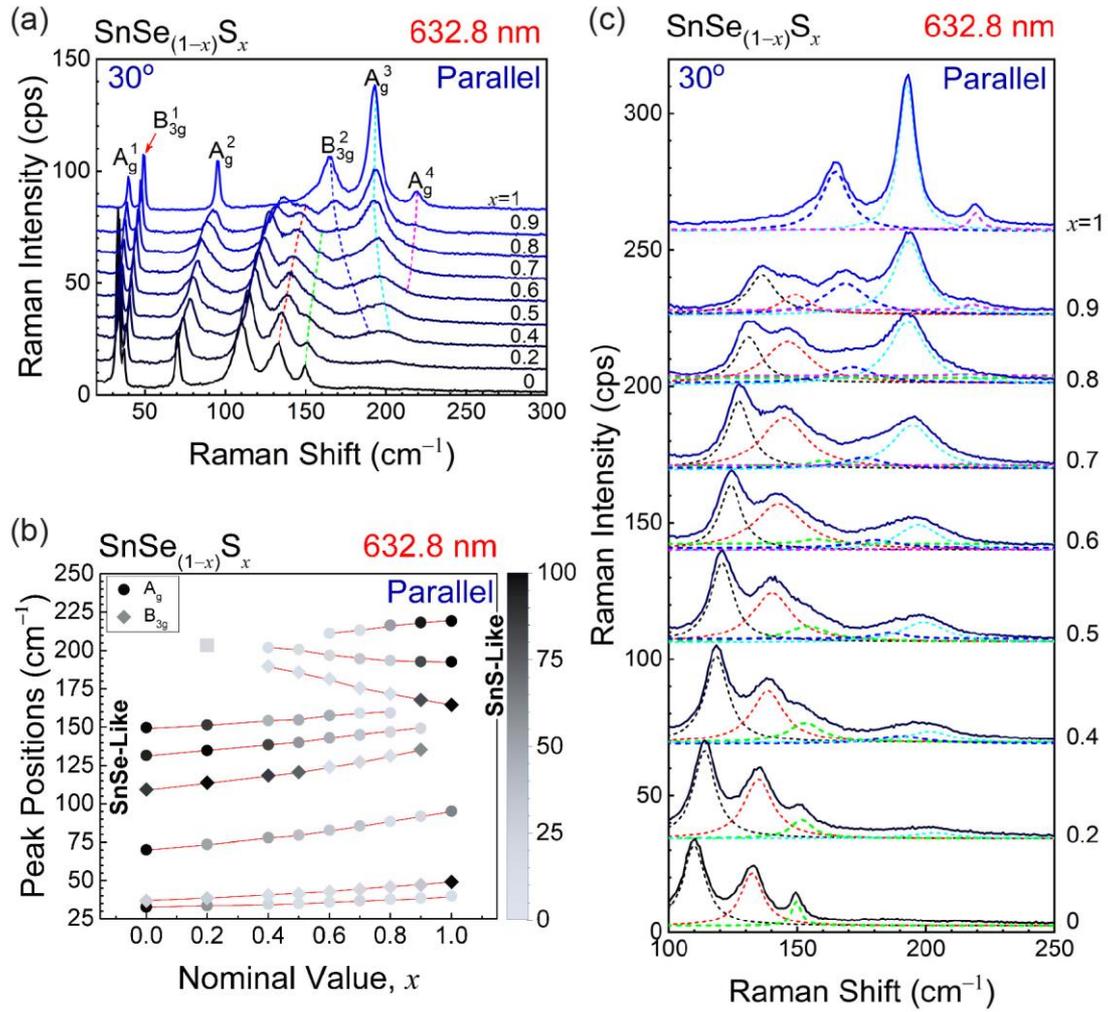

**Figure 3.** (**a**) Raman spectra of SnSe$_{(1-x)}$S$_x$ ($0 \leq x \leq 1$) single crystals measured by using the 632.8-nm excitation wavelength in parallel polarization configuration at 30° with respect to the armchair direction. Dotted red lines are guide for the eyes. (**b**) Raman peak positions as a function of nominal composition ($x$). The intensity of each mode is normalized to the maximum intensity of the mode in the entire composition range. The gray square at ~203 cm$^{-1}$ for $x$=0.2 indicates the overlapped peaks of SnS-like B$_{3g}^2$ and A$_g^3$. (**c**) Enlarged Raman spectra of SnSe$_{(1-x)}$S$_x$ ($0 \leq x \leq 1$) single crystals from (a). The spectra are deconvoluted using Lorentzian functions to show individual Raman modes in the alloys.

The general trend of the blue shift of the peaks as the S content ($x$) is increased can be understood in terms of the decreasing average reduced mass as lighter S is substituted for Se combined with the changed bond strength due to the smaller size of S atoms as reviewed by Chang and Mitra[36]. A similar effect was seen for the low-frequency shear mode in $Mo_xW_{(1-x)}S_2$[37]. However, the opposite trends of the SnS-like $B_{3g}^2$ and $A_g^3$ modes are intriguing and cannot be explained in such a simple model. It is suspected that the Se substitution for S somehow modifies the charge distribution around the S atoms in such a way to increase the Sn-S bonding strength. Similar trends have been reported for $Mo_{1-x}W_xS_2$ and $Ge_xSi_{1-x}$[38-39]. More detailed theoretical analysis, which is beyond the scope of this work, would be needed to understand this peculiarity.

The co-existence of one- and two-mode behavior in mixed crystals was also observed in the $GaS_xSe_{1-x}$, $KMgF_3$-$KNiF_3$ and in some other materials discussed by Chang and Mitra[40-41, 36]. In several mixed crystal systems, one can immediately recognize one-mode type by observing the peaks being continuously evolved from one end to the other of the alloy compositions, whereas disconnection is found in two-mode type. The different behavior can be understood in terms of the actual vibrations of the atoms in each mode. For example, the $B_{3g}^1$ mode that shows the one-mode behavior is a vibration of the entire layer against the neighboring layer (see Figure 2(f)). No individual Sn-S or Sn-Se bonds are modified by this mode. When S-Se substitution is made, the average mass of the layers varies in proportion to the atomic ratio. Therefore, the normal mode frequency should vary continuously as a function of $x$. In contrast, the $A_g^3$ mode, which shows the two-mode behavior, involves stretching vibrations of the Sn-S or Sn-Se bonds along the $x$ axis (Figure 2(f)). Since individual bonds are modified by the vibration, substitution of the chalcogen atoms results in coupled oscillations of Sn-S and Sn-Se vibrations, which leads to two vibration modes. For example, when S is substituted for Se in pure SnSe, the SnSe vibration mode is slightly modified, whereas a local vibration mode for the Sn-S bonds appears at higher frequency due to the smaller mass of S. An interesting case is the $A_g^2$ mode. If the evolution of the Raman spectrum is examined as a function of the composition (Figure 3(a)), the $A_g^2$ peak, which is as sharp as the other lower frequency peaks for the pure compounds SnSe and SnS, is significantly broadened in the alloy case, whereas the widths of the $A_g^1$ and $B_{3g}^1$ peaks do not show such alloy broadening (see Supplementary Figure S12). Similar broadening of phonons with one-mode behavior has been observed in the E2 mode of the $Al_xGa_{1-x}N$ films and has been ascribed to the disorder due to the composition fluctuation in the alloy[42]. We note that the $A_g^1$ and $B_{3g}^1$ modes show much less broadening. This is because these modes represents vibrations of the entire layer as a rigid body against another layer, and so the compositional fluctuation within a layer does not affect the overall vibration frequency significantly other than the effect of the change in the mass of the layer that gives the shift with the composition.

In order to determine the crystal axes of the samples, TEM analysis was performed on several SnSe, SnS and $SnSe_{0.5}S_{0.5}$ samples. Figure 4(a-b) and Figure 4(h-i) are TEM images of SnSe and SnS samples, respectively. Figure 4(c) and Figure 4(j) are the corresponding selected area electron diffraction (SAED) patterns, from which the armchair (AC) and zigzag (ZZ) directions are determined (see Supplementary Figure S13 for the data on $SnSe_{0.5}S_{0.5}$). Figure 4(d-g) and Figure 4(k-n) are the polar plots of the intensities of $A_g$ modes measured on the samples on TEM grids with three different excitation wavelengths as indicated. The orientation of the samples were carefully matched between the TEM and Raman measurements. Some of the modes show dramatically different polarization dependences for different excitation wavelengths: for example, the maximum intensity direction of the $A_g^4$ mode of SnSe changes with the excitation wavelength. We find that the polarization dependence of the $A_g^2$ mode varies the least with the excitation wavelength, and the maximum intensity direction of this mode corresponds to the armchair direction.

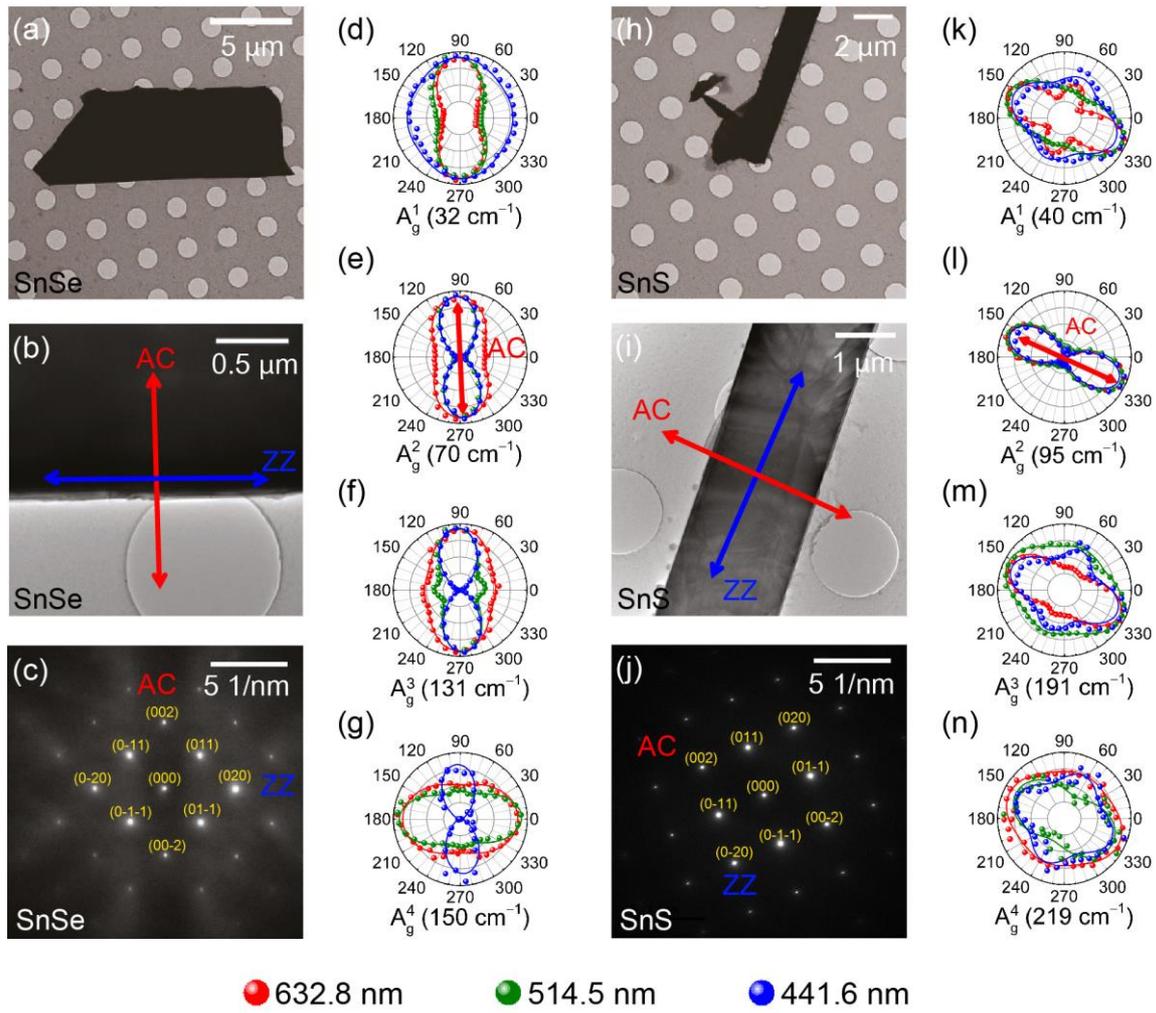

**Figure 4**. Comparison of transmission electron microscopy (TEM) and polarized Raman results measured in parallel polarization configuration to determine crystallographic directions of SnSe and SnS. (**a,b,h,i**) and (**c,j**) are TEM images and the selected area electron diffraction (SAED) patterns of SnSe and SnS samples, respectively. The red and blue arrows in (**b**) and (**i**) indicate the armchair (AC) and zigzag (ZZ) directions of SnSe and SnS, respectively. (**d-g**) and (**k-n**) are polarization dependence of the integrated intensity of the $A_g$ modes of SnSe and SnS samples, respectively. The maximum intensity is normalized for each polar plot to show the polarization dependence clearly. The curves represent best fits to the calculated polarization dependence of the Raman intensities using equation (1). The 632.8, 514.5 and 441.6-nm excitation wavelengths are used in the measurements as indicated.

For more comprehensive information, Figure 5 compares the polarization dependences of the Raman modes in SnSe, SnSe$_{0.5}$S$_{0.5}$, and SnS measured with six excitation wavelengths. The armchair direction is taken as 0°. As expected from equation (2), the polarization dependence of the B$_{3g}$ mode does not vary with the excitation wavelength. For the A$_g$ modes except the A$_g^2$ mode, the maximum intensity direction changes with the excitation wavelength even for the same composition. For example, for SnSe, the A$_g^4$ mode is maximum along the armchair direction when the 441.6 nm excitation is used but is along the zigzag direction for the other excitation wavelengths (see Figure 5(e)). In addition, the polarization dependences of the Raman peaks in anisotropic 2D materials sometimes vary significantly with the thickness of the sample due to the interference effect combined with the birefringence and linear dichroism of the crystal[17, 43]. As one can see from the Supplementary Figures S14 and S15, the A$_g^2$ mode varies little with the sample thickness, whereas the other A$_g$ modes vary significantly. As explained earlier, birefringence, linear dichroism, the anisotropy of the electron-phonon interaction are believed to be responsible for this peculiar dependence of the polarization behavior on the excitation laser wavelength. Here, one should note that most of the laser energies used were not close to the resonance at ~1.58 eV and at ~2.4 eV, and so the resonance effect is not likely to contribute significantly to this phenomenon. We also measured the Raman spectra as a function of the excitation laser intensity and found no significant (see Supplementary Figure S16). Therefore, one should use the polarization dependence of the A$_g^2$ mode for reliable determination of the crystallographic direction regardless of the thickness.

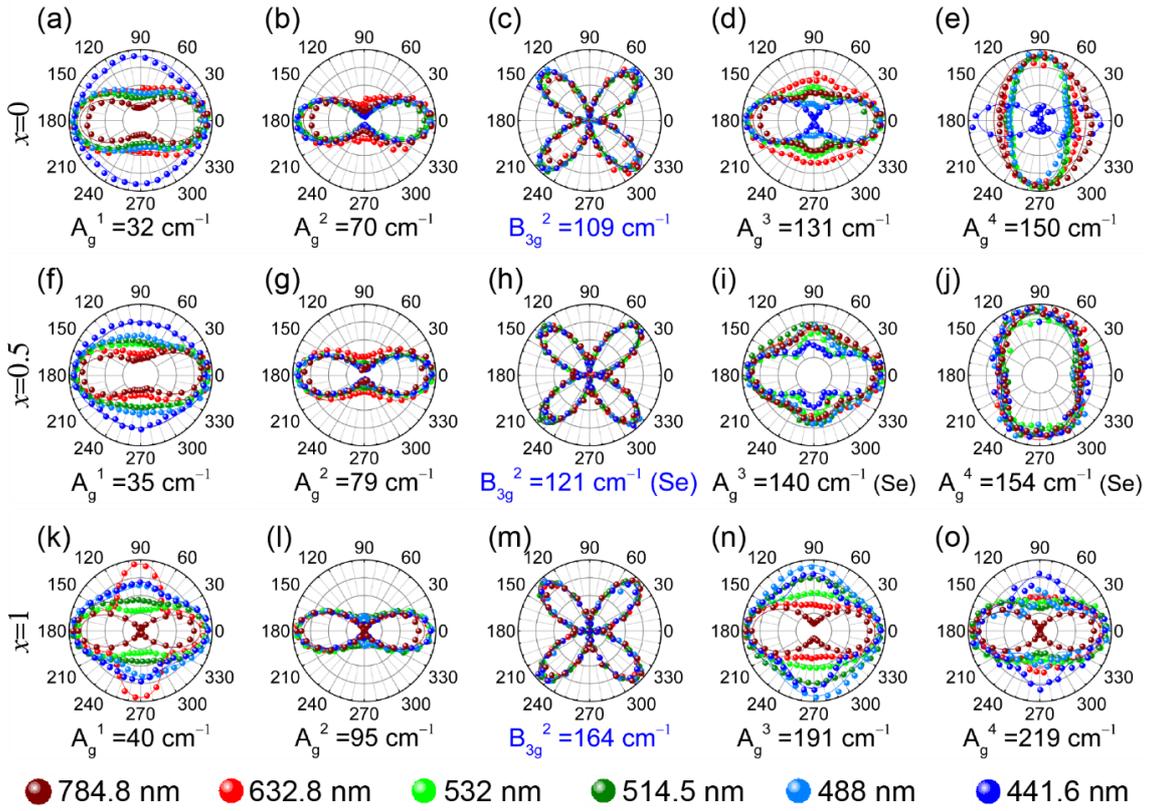

**Figure 5**. Polarization dependence of integrated intensity of the main modes in SnSe$_{(1-x)}$S$_x$ ($x$=0, 0.5, 1) single crystals measured in parallel polarization configuration and by using six different excitation wavelengths as indicated. The Se in the parentheses in SnSe$_{0.5}$S$_{0.5}$ ($x$=0.5) indicates that the mode is an SnSe-like mode. The maximum intensity is normalized for each polar plot to show the polarization dependence clearly. The curves represent best fits to the calculated polarization dependence of the Raman intensities using equation (1) and (2).

To sum up, detailed polarized Raman spectroscopic measurements were carried out on SnSe$_{(1-x)}$S$_x$ alloys to establish the correlation between the polarization dependence of the Raman modes with the in-plane crystallographic directions. It was found that the polarization behavior of some of the Raman modes vary significantly with the excitation laser wavelengths used and that the 632.8-nm excitation wavelength is the most suitable laser for such purposes. By comparing the polarized Raman studies with electron microscopy, it is established that the maximum intensity direction of the $A_g^2$ mode measured in the parallel polarization configuration corresponds to the armchair direction. In the composition dependence of the Raman spectrum, it is observed the one-mode behavior for the lower-frequency layer-by-layer vibration modes and the two-mode behavior for the high-frequency modes that involve local Sn-S(e) vibrations. The results provide an easy and foolproof method to determine the crystallographic direction of $SnSe_{(1-x)}S_x$, which is an important parameter in fabricating devices using this material. Furthermore, making unambiguous phonon mode assignments and establishing their evolution as a function of the composition will benefit the analysis of other compounds such as the solar cell materials $Cu_2ZnSn(S,Se)_4$ in which Sn(S,Se) are often found as a secondary phase material that limit the cell performance.

**Methods**

The $SnSe_{(1-x)}S_x$ single crystals ($x$=0, 0.2, 0.4, 0.5, 0.6, 0.7, 0.8, 0.9, 1) were grown by using the temperature gradient method[27-29]. In this method, tin (99.8%), selenium (99.9%), and sulfur (99%) powders were first weighed at a molecular ratio of 1:(1–$x$):$x$, respectively. They were then loaded into quartz ampoules before being pumped down (<$10^{-4}$ Torr) and sealed. The sealed ampoules were loaded into a vertical furnace and gradually heated up to 960 °C then soaked at that temperature for 33 hours in order to homogenize the melt. In the final process, the temperature was cooled down at a speed of 1 °C h$^{-1}$ from 960 °C to 600 °C and then at a relatively higher speed of 20 °C h$^{-1}$ to room temperature. The single crystals were finally obtained after breaking the ampoules. The crystallographic structures of the crystals were examined by x-ray diffraction (XRD). The chemical compositions of the crystals were determined by comparing the XRD data with the energy dispersive x-ray spectroscopy (EDS). The bulk crystals cleave parallel to the (100) plane, and Raman measurements were performed on cleaved surfaces. The same measurements were repeated on samples exfoliated on $SiO_2$/Si substrates of thickness in the range of ~50 to ~300 nm as determined by using a commercial atomic force microscope (AFM) system (NT-MDT NTEGRA Spectra) and the results were identical in terms of Raman peak positions. For polarized Raman measurements, a low-frequency polarized Raman setup was used in the backscattering geometry[32-33] with six different excitation wavelengths: the 784.8-nm (1.58 eV) line of a diode laser, the 632.8-nm (1.96 eV) line of a He-Ne laser, the 532-nm (2.33 eV) line of a diode-pumped solid-state (DPSS) laser, the 514.5-nm (2.41 eV) and 488-nm (2.54 eV) lines of an Ar ion laser and the 441.6-nm (2.81 eV) line of a He-Cd laser. The laser beam is linearly polarized and focused onto the sample by a 50× microscope objective lens (N.A. 0.8). The laser power was maintained below 300 μW before the objective lens in order to minimize local heating and photochemical reactions. A polarizer is placed after the laser source to make a linearly polarized beam. A half-waveplate was placed before the objective lens to control the incident polarization. The scattered beam was collected and collimated by the same objective lens, and an analyzer was used to select the polarization of the scattered signal. An additional half-waveplate was used right before the focusing lens in front of the entrance slit of the spectrometer to make the polarization of the light entering the spectrometer constant regardless of the analyzer direction. A set of three volume holographic filters (Ondax for 784.8-nm line and OptiGrate for the others) were used to observe the Raman modes of the alloys in the low-frequency range below 100 $cm^{-1}$. The Raman scattering signals were dispersed by a Jobin-Yvon iHR550 spectrometer with a 1200 grooves/mm grating (630 nm blaze) for the 632.8-nm and 784.8-nm lines or a 2400 grooves/mm grating (400 nm blaze) for the other laser lines to get the optimal resolution and the signal-to-noise ratio. The signal was detected by using a liquid-nitrogen-cooled back-illuminated charged-couple-device (CCD) detector (Horiba Symphony II BIDD). The spectral resolution was below 1 $cm^{-1}$.

Several mechanically exfoliated SnSe, SnS and SnSe$_{0.5}$S$_{0.5}$ samples were used in transmission electron microscopy (TEM) for determination of zigzag and armchair directions in order to match with the results of polarized Raman spectroscopy. Thin poly-methyl-methacrylate (PMMA) was first spin coated at a speed of 6000 revolutions per minute (RPM) onto 300-nm SiO$_2$/Si substrates. Several bulk SnSe, SnS and SnSe$_{0.5}$S$_{0.5}$ samples were then made on the substrates by using mechanical exfoliation. The resulting samples were further identified by using optical microscopy (Leica DM-750M). A carbon film coated Quantifoil TEM grid was attached onto the substrate and the exfoliated samples were then transferred onto the TEM grid through the removal of the PMMA layer in acetone. TEM imaging, selected area electron diffraction (SAED) and energy dispersive x-ray spectroscopy (EDS) of TEM samples were performed with a double Cs-aberration corrected JEOL ARM-200F operated at 80 kV.

**Acknowledgements**

This work was supported by the Korea Institute of Energy Technology Evaluation and Planning (KETEP) and the Ministry of Trade, Industry & Energy (MOTIE) of the Republic of Korea (No. 20173010012980) and the Basic Science Research Program through the National Research Foundation of Korea (NRF-2017R1A5A1014862, NRF-2019R1F1A1058473 and NRF-2019R1A6A1A11053838) and Institute for Basic Science (IBS-R026-D1). T.S. acknowledges the supports from Swedish International Development Cooperation Agency (SIDA) through Sweden and Royal University of Phnom Penh (RUPP)'s Pilot Research Cooperation Programme (Sida Contribution No. 11599) and RUPP.

**Author contribution statements**

H.C. conceived the experiments. T.S. and S.Y.L. carried out the Raman experiments. T.M.H.N., V.Q.N., and S.C. grew the single crystals, and Y.L. and K.K. performed transmission electron microscopy (TEM). The data analyses were done by T.S. and H.C.

All authors have contributed to the writing of the manuscript and have given approval to the final version.

## Additional information

**Supplementary Information** accompanies the paper at http://www.nature.com/srep.

## Declaration of competing interest

The authors declare no competing interests.